\documentclass[twocolumn,aps,pra,floatfix,superscriptaddress,notitlepage]{revtex4-1}
\usepackage[utf8]{inputenc}
\usepackage[OT1]{fontenc}
\usepackage{amsmath}
\usepackage{amsfonts}
\usepackage{amssymb}
\usepackage{bm}
\usepackage{graphicx}
\usepackage[left=2cm,right=2cm,top=2cm,bottom=2cm]{geometry}
\usepackage{longtable,booktabs}
\allowdisplaybreaks

\usepackage{dcolumn}
\usepackage{siunitx}
\usepackage{multirow}
\usepackage{bigdelim}

\usepackage{color}
\definecolor{BLUE}{rgb}{0.0,0.0,1.0}

\usepackage{soul}
\soulregister\cite7
\soulregister\ref7

\newcommand{\veps}{\varepsilon}
\newcommand{\balpha}{\bm{\alpha}}
\newcommand{\bnabla}{\bm{\nabla}}

\newcommand{\br}{\bm{r}}

\newcommand{\bp}{\bm{p}}

\newcommand{\be}{\begin{eqnarray}}
\newcommand{\ee}{\end{eqnarray}}

\begin{document}

\title{\textit{Ab initio} calculations of the $2p_{3/2} \rightarrow 2s$ transition in He-, Li-, and Be-like uranium}

\author{A.~V.~Malyshev}
\affiliation{Department of Physics, St.~Petersburg State University, Universitetskaya 7/9, 199034 St.~Petersburg, Russia  }

\author{Y.~S.~Kozhedub}
\affiliation{Department of Physics, St.~Petersburg State University, Universitetskaya 7/9, 199034 St.~Petersburg, Russia  }

\author{V.~M.~Shabaev}
\affiliation{Department of Physics, St.~Petersburg State University, Universitetskaya 7/9, 199034 St.~Petersburg, Russia  }
\affiliation{Petersburg Nuclear Physics Institute named by B.P. Konstantinov of National Research Center ``Kurchatov Institute'', Orlova roscha 1, 188300 Gatchina, Leningrad region, Russia}


\begin{abstract}

The bound-state QED approach is applied to calculations of the $2p_{3/2} \rightarrow 2s$ transition energies in He-, Li-, and Be-like uranium. For U$^{90+}$ and U$^{89+}$, standard perturbation theory for a single level is employed, while the calculations of U$^{88+}$ have required its counterpart for quasidegenerate levels. The utilized approach merges the rigorous QED treatment up to the second order of perturbation theory with the higher-order electron-correlation contributions evaluated within the Breit approximation. The higher-order screened QED effects are estimated by means of the model-QED operator. The nuclear recoil, nuclear polarization, and nuclear deformation effects are taken into account as well. Along with the transition energies, their pairwise differences are calculated. The comprehensive analysis of the uncertainties due to uncalculated effects is carried out, and the most accurate theoretical predictions, which are in perfect agreement with available experimental data, are obtained.

\end{abstract}


\maketitle


\section{Introduction \label{sec:0}}

Highly charged ions are generally recognized as an excellent testing bed to examine high-precision methods of bound-state quantum electrodynamics (QED) in the presence of strong electromagnetic fields~\cite{Sapirstein:2008:25, Beiersdorfer:2010:074032, Glazov:2011:71, Volotka:2013:636, Shabaev:2018:60, Kozlov:2018:045005, Indelicato:2019:232001}. This is because a fairly small number of electrons in these systems enables an accurate treatment of the electron-electron correlations not masking the QED effects substantially enhanced compared to light atoms. Along with this, the nuclear-strength parameter~$\alpha Z$ ($\alpha$ is the fine-structure constant and $Z$ is the atomic number), being close to unity, makes the $\alpha Z$-expansion approaches~\cite{Caswell:1986:437} inapplicable, which leads to the need for carrying out all the calculations nonperturbatively. This feature mainly distinguishes the highly charged ions from the low-$Z$ systems that historically were the focus of extensive theoretical and experimental investigations.

If restricted to the discussion of the electronic structure, then the most stringent test of the bound-state QED methods at the strong-coupling regime is provided currently by the ground-state Lamb-shift measurement in H-like uranium~\cite{Stoehlker:2000:3109, Gumberidze:2005:223001} and the precise determination of the $2p_{1/2} \rightarrow 2s$ transition in Li-like uranium~\cite{Schweppe:1991:1434, Brandau:2003:073202, Beiersdorfer:2005:233003}. The related theory can be found, e.g., in Refs.~\cite{Yerokhin:2006:253004, Kozhedub:2008:032501, Sapirstein:2011:012504}. Other charge states of uranium and other transitions also represent the object of the experimental studies~\cite{Munger:1986:2927, Briand:1990:2761, Beiersdorfer:1993:3939, Beiersdorfer:1995:114, Beyer:1995:169, Beiersdorfer:1998:1944, Gumberidze:2004:203004, Beiersdorfer:2005:233003, Trassinelli:2009:63001, Trassinelli:2009:012026, Trassinelli:2011:014003, Nakano:2013:060501_R}. In particular, in Refs.~\cite{Trassinelli:2009:63001, Trassinelli:2011:014003} the first clear identification and measurement of the $2p_{3/2} \rightarrow 2s$ transition in He-like uranium was reported. The energy difference between the $2p_{3/2} \rightarrow 2s$ transitions in He- and Li-like ions was evaluated as well. Both presented values were found in perfect agreement with the available at that time theoretical predictions. Since then, much effort has been put into this issue from the experimental side, and a considerable improvement in the accuracy as well as the extension of the analysis to the case of Be-like uranium are anticipated in the near future~\cite{FAIR_proposal_E125, Trassinelli:private}. It is announced that the accuracy of the new experiment, conducted at the ESR storage ring at the GSI in Darmstadt, is more than one order of magnitude higher than in the previous measurement~\cite{GSI_report_2021}. With this in mind, the main goal of the present work is to perform the benchmark \textit{ab initio} calculations of the $2p_{3/2} \rightarrow 2s$ transition energies and all possible energy differences between the corresponding lines in He-, Li-, and Be-like uranium.

In the literature, there are numerous relativistic calculations of the $2p_{3/2} \rightarrow 2s$ transitions in few-electron uranium ions~\cite{Cheng:1979:111, Drake:1988:586, Kim:1991:148, Chen:1993:3692, Chen:1995:266, Johnson:1995:297, Plante:1994:3519, Safronova:1996:4036, Cheng:1996:2206, Chen:1997:166, Santos:1998:149, Cheng:2000:054501, Artemyev:2005:062104, Yerokhin:2007:062501, Cheng:2008:052504, Kozhedub:2010:042513, Sapirstein:2011:012504, Kaygorodov:2019:032505, Kozhedub:2019:062506}. State-of-the-art QED methods to treat the energy levels in highly charged ions comprise all the contributions up to the second order of perturbation theory that implies the rigorous consideration of the one-electron one- and two-loop and two-electron one-loop QED effects as well as the contributions of the two-photon-exchange diagrams. Such calculations for He- and Li-like ions have been performed, e.g., in Refs.~\cite{Artemyev:2005:062104} and \cite{Yerokhin:2007:062501, Kozhedub:2010:042513, Sapirstein:2011:012504}, respectively. Recently, we have reevaluated the $n=1$ and $n=2$ energy levels for a number of He-like ions including uranium~\cite{Malyshev:2019:010501_R, Kozhedub:2019:062506}. In contrast to these higher charge states, all the available theoretical predictions for the $2p_{3/2} \rightarrow 2s$ transition in Be-like uranium were obtained taking into account the QED effects at best within some first-order one-electron approximations. In Refs.~\cite{Malyshev:2021:183001, Malyshev:2021:652}, we have shown that, because of the strong level mixing, the proper description of the electronic structure of Be-like ions can be given only within the framework of QED perturbation theory for quasidegenerate levels. In the present work, the $2p_{3/2} \rightarrow 2s$ transition in Be-like uranium will be addressed by means of the corresponding most advanced methods~\cite{Malyshev:2019:010501_R, Kozhedub:2019:062506, Malyshev:2021:183001, Malyshev:2021:652}. The related transitions in He- and Li-like uranium ions will be revisited within the similar approaches to provide a consistent QED treatment for all three charge states of uranium and to obtain the most accurate values of the pairwise transition-energy differences. A brief description of the employed methods is presented in the next section. The discussion of the numerical results is given in Sec.~\ref{sec:2}.

Relativistic units ($\hbar=1$ and $c=1$) and Heaviside charge unit ($e^2=4\pi\alpha$, where $e<0$ is the electron charge) are used throughout the paper.


\section{Theoretical approach and computational details \label{sec:1}}

Let us briefly describe the approach utilized in the present work for the evaluation of the $2p_{3/2} \rightarrow 2s$ transition in He-, Li-, and Be-like uranium. The zeroth-order approximation is given by the Dirac equation
\begin{equation}
\label{eq:DirEq}
h^{\rm D}\psi_n \equiv \left[ \balpha \cdot \bp + \beta m + V \right] \psi_n = \veps_n \psi_n \, ,
\end{equation}
where $\bp=-i\bnabla$ is the momentum operator, and $\balpha$ and $\beta$ are the Dirac matrices. We use several variants of the spherically-symmetric binding potential~$V$ in Eq.~(\ref{eq:DirEq}) which lead to different realizations of the QED perturbation series. The simplest choice is the Coulomb (Coul) potential of the nucleus, $V=V_{\rm nucl}$. This option corresponds to the Furry picture of QED~\cite{Furry:1951:115} and ensures that the electron-nucleus interaction is treated nonperturbatively (to all orders in $\alpha Z$). We also employ the variants with some local screening potentials added to the nuclear potential, $V=V_{\rm nucl} + V_{\rm scr}$. The inclusion of the screening potential into the initial approximation defines the so-called extended version of the Furry picture and serves to take into account along with the electron-nucleus coupling some part of the interelectronic interaction from the very beginning. If it were possible to calculate the QED contributions to all orders of perturbation theory, the final results would be independent of the choice of the zeroth-order Hamiltonian. Therefore, the comparison of the calculations with the different types of the screening potentials allows us to judge the convergence of the results and partly catch the higher-order QED effects. 

In the calculations, we use four different screening potentials. The first and the second choices are the core-Hartree potentials generated by one and two $1s$ electrons, respectively. We will refer to these potentials as CH1 and CH2. The CH1 potential was employed previously in Ref.~\cite{Kozhedub:2019:062506} to study He-like uranium. The third potential is the local Dirac-Fock (LDF) potential~\cite{Shabaev:2005:062105}, which is constructed from the wave functions obtained within the Dirac-Fock approximation for the configuration~$1s^2 2s$. Finally, the last potential is the Kohn-Sham (KS) potential~\cite{pot:KS} corresponding to the configuration~$1s^2 2s^2$. All the potentials are determined self-consistently. We note that the asymptotic behavior of the screening potentials at large distances~$r$ has the form of $(\alpha N_{\rm scr})/r$, where $N_{\rm scr}=1$ for CH1, $N_{\rm scr}=2$ for CH2 and LDF, and $N_{\rm scr}=3$ for KS. The Latter correction~\cite{Latter:1955:510} is introduced to restore the proper asymptotic behavior of the KS potential. 

In all the calculations, the finite-nuclear-size (FNS) effect is taken into account by considering the nuclear potential $V_{\rm nucl}$ for the extended nucleus. The two-parameter Fermi model is employed to describe the nuclear-charge distribution,
\begin{equation}
\label{eq:fermi}
\rho_{\rm nucl}(r) = \frac{\rho_0}{1+\exp\left[\left(r-c\right)/a\right]} \, ,
\end{equation}
where the parameter~$a$ is fixed by the standard choice of $a=2.3/(4\ln3)$~fm, the parameter~$c$ is related to the root-mean-square (RMS) radius~$R$ by the approximate formula
\begin{equation}
\label{eq:c}
c^2 = \frac{5}{3} R^2 - \frac{7}{3} a^2 \pi^2 \, ,
\end{equation}
and $\rho_0$ is the normalization factor. For the aims of the present work, the accuracy of Eq.~(\ref{eq:c}) is sufficient. The value of $R$ is taken for $^{238}$U from Refs.~\cite{Angeli:2013:69, Yerokhin:2015:033103}. The Dirac Eq.~(\ref{eq:DirEq}) is solved in the basis of $B$ splines~\cite{Johnson:1988:307, Sapirstein:1996:5213} within the dual-kinetic-balance approach~\cite{splines:DKB}. We employ the obtained finite-basis set in the subsequent QED calculations. In order to reduce the uncertainty related to the FNS effect, the nuclear deformation correction is explicitly taken into account for the Dirac energies as calculated in Ref.~\cite{Kozhedub:2008:032501}. It has been reevaluated with respect to the used values of the Fermi-model parameters and constitutes $-0.10(19)$~eV, $-0.020(34)$~eV, and $-0.002(4)$~eV for the $1s$, $2s$, and $2p_{1/2}$ states, correspondingly. The values of the fundamental constants from Ref.~\cite{Tiesinga:2021:025010} are employed throughout the calculations.

Within the zeroth-order approximation, the many-electron energies are given by the sums of the Dirac eigenvalues, while the unperturbed wave functions are constructed from the corresponding bispinors in the $jj$ coupling. The effects due to the interaction with the quantized electromagnetic field and electron-electron interaction are treated by means of QED perturbation theory (PT), which we formulate in the framework of the two-time Green's function (TTGF) method~\cite{TTGF}. For He- and Li-like uranium, the states of interest are well isolated from other levels of the same symmetry. Therefore, standard PT for a single level suits for their accurate description. As we have convincingly demonstrated in Refs.~\cite{Malyshev:2021:183001, Malyshev:2021:652}, this is not the case for Be-like ions. QED PT for quasidegenerate levels has to be used instead. This PT implies the construction of the effective Hamiltonian~$H$ acting in the model subspace~$\Omega$ spanned by  close levels under consideration, and the desired energies are the eigenvalues of~$H$, see Refs.~\cite{TTGF} for details. In Ref.~\cite{Malyshev:2021:183001}, we studied the ground and low-lying excited states in Be-like xenon. We have shown that the accurate treatment of the QED effects for the ground state can be achieved if at least the $1s^2 2s^2$, $1s^2 2p_{1/2}^2$, and $1s^2 (2p_{3/2}2p_{3/2})_0$ states are included into the three-dimensional model subspace~$\Omega$, and all of them are considered within PT for quasidegenerate levels. Furthermore, the states with the total angular momentum~$J$ equal to 1, namely $1s^2 (2s2p_{1/2})_1$ and $1s^2 (2s2p_{3/2})_1$, also have to be combined into the model subspace and calculated together. For Be-like uranium, however, it turns out that the $2p_{3/2} \rightarrow 2s$ transition can be evaluated with the required precision, if one treats the $1s^2 (2s2p_{3/2})_1$ level as isolated. Moreover, when considering the ground state, it is sufficient to form the two-dimensional model subspace~$\Omega$ by taking only the $1s^2 2s^2$ and $1s^2 2p_{1/2}^2$ levels, while the inclusion of the $1s^2 (2p_{3/2}2p_{3/2})_0$ state in~$\Omega$ is not as crucial as for lighter Be-like ions. Nevertheless, since the present work focuses not only on the calculations of the transition energies, but also on their differences, for which a higher accuracy is anticipated, both ground and excited states are considered within PT for quasidegenerate levels, and, in particular, the three-dimensional subspace~$\Omega$ is constructed to determine the ground-state energy. In Sec.~\ref{sec:2}, despite the fact that the applied formalism is fully relativistic, we employ the $LS$-coupling notations instead of the $jj$-coupling ones to designate the investigated states. For He- and Li-like ions, the correspondence between two couplings is unambiguous: the states with the maximum values of $J$ are considered for U$^{90+}$, while the electronic configurations with one valence electron over the closed $1s^2$ shell are dealt in the case of U$^{89+}$. In a sense, this is a rationale for the use of PT for a single level. On the contrary, for Be-like uranium, the $LS$ notations emphasize once again that the considered states actually differ from the levels resulting from the standard single-level perturbative QED approach and are obtained by diagonalizing the corresponding matrices~$H$.

\begin{figure}
\begin{center}
\includegraphics[width=\columnwidth]{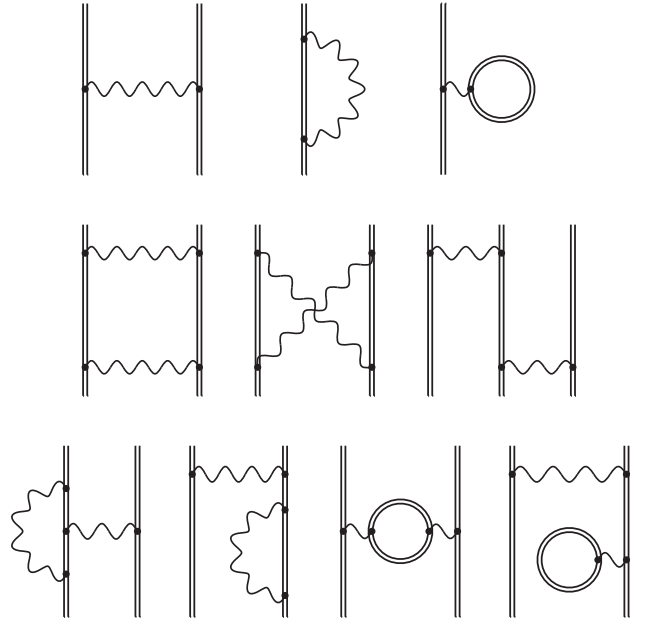}
\caption{\label{fig:diagr}
First- and second-order Feynman diagrams the contributions of which are rigorously evaluated in the present work (in each row from left to right): one-photon-exchange diagram, one-electron self-energy and vacuum-polarization diagrams (first line); two-photon-exchange diagrams (second line); two-electron self-energy and vacuum-polarization diagrams (third line). The counterterm diagrams are not shown. The double line denotes the electron propagator in the binding potential~$V$. The wavy line corresponds to the photon propagator.}
\end{center}
\end{figure}

The details of the QED calculations are discussed in our previous works, see, e.g., Refs.~\cite{Malyshev:2014:062517, Malyshev:2017:022512, Kozhedub:2019:062506, Malyshev:2019:010501_R, Malyshev:2021:183001, Malyshev:2021:652}. Regardless of the PT type, that is, both for a single level and a set of quasidegenerate levels, we consider all the QED contributions up to the second order. Namely, we rigorously evaluate the contributions of the one- and two-photon-exchange and one- and two-electron self-energy and vacuum-polarization diagrams shown in Fig.~\ref{fig:diagr}. In the case of the extended Furry picture, the contributions of the related counterterm diagrams are calculated as well. All computational formulas are derived within the TTGF method. For Be-like ions, the formalism with the closed $K$ shell assigned to the vacuum~\cite{TTGF} has been employed to obtain the formal expressions. The one-electron two-loop QED corrections for the Coulomb potential are taken into account employing the results from Refs.~\cite{Yerokhin:2015:033103, Yerokhin:2018:052509}. For the screening potentials, these Coulomb values are scaled using the first-order QED corrections for the $ns$ electrons, where $n$ is the principal quantum number. 

The remaining interelectronic-interaction effects of third and higher orders are taken into account within the Breit approximation using the Dirac-Coulomb-Breit (DCB) Hamiltonian~\cite{Faustov:1970:478, Sucher:1980:348, Mittleman:1981:1167},
\begin{align}
\label{eq:DCB}
&H_{\rm DCB} = \Lambda^{(+)} \left[ H_0 + V_{\rm int} \right] \Lambda^{(+)} \, , \\
\label{eq:H0}
&H_0 = \sum_i \left[ \balpha_i \cdot \bp_i + \beta_i m + V_{\rm nucl}(r_i) \right] \, , \\ 
\label{eq:Vint}
&V_{\rm int} = \sum_{i<j} \left[ V^{\rm C}_{ij} + V^{\rm B}_{ij} \right] \, , 
\end{align}
where the indices $i$ and $j$ enumerate the electrons, and the Coulomb and Breit parts of the interelectronic-interaction operator~$V_{\rm int}$ read as
\begin{align}
\label{eq:V_C}
V^{\rm C}_{ij} &= \frac{\alpha}{r_{ij}} \, , \\ 
\label{eq:V_B}
V^{\rm B}_{ij} &= -\frac{\alpha}{2} \left[ \frac{\balpha_i \cdot \balpha_j}{r_{ij}}
+ \frac{(\balpha_i \cdot \br_{ij})(\balpha_j \cdot \br_{ij})}{r_{ij}^3}  \right] \, .
\end{align}
The operator~$\Lambda^{(+)}$ in Eq.~(\ref{eq:DCB}) is the projector on the states constructed from the positive-energy eigenfunctions of the one-electron Dirac Hamiltonian~$h^{\rm D}$, which determines the initial approximation for the QED calculations and is defined by Eq.~(\ref{eq:DirEq}). In principle, in relativistic atomic calculations, other one-electron operators, e.g., nonlocal Dirac-Fock operator~$h^{\rm DF}$, are often used to determine the projectors~$\Lambda^{(+)}$. However, our choice naturally arises from QED~\cite{Shabaev:1993:4703} and provides the most consistent merging with the QED calculations, see, e.g., the related discussion in Ref.~\cite{Kozhedub:2019:062506}. The DCB equation is solved by the configuration-interaction (CI) method in the basis of the Dirac-Sturm orbitals~\cite{Bratzev:1977:173, Tupitsyn:2003:022511}. In the case of PT for a single level, the procedure to separate the contribution of third and higher orders from the total CI energies is straightforward, see, e.g., Refs.~\cite{Kozhedub:2010:042513, Malyshev:2017:022512}. For quasidegenerate levels, the corresponding procedure is more intricate, and it was proposed and realized for the first time in Ref.~\cite{Malyshev:2019:010501_R}, where the $n=2$ states in middle-$Z$ He-like ions were studied. The main idea of the method consists in adding the higher-order correlation effects to the matrix of the effective Hamiltonian~$H$, which describes the quasidegenerate levels, before its diagonalization. The detailed description of the procedure can be found in Refs.~\cite{Kozhedub:2019:062506, Malyshev:2021:652}. In the case of the extended Furry picture, it is convenient to rearrange the expression in the square brackets in Eq.~(\ref{eq:DCB}) so that the orders of PT still correspond to the powers of the modified operator~$V_{\rm int}$. Since the modified unperturbed part~$H_0$ is obtained  from Eq.~(\ref{eq:H0}) by adding the screening potential~$V_{\rm scr}$ for each electron, $H_0=\sum_i h_i^{\rm D}$, one readily sees that the modified interaction operator is given by
\begin{align}
\label{eq:Vint_scr}
V_{\rm int} = \sum_{i<j} \left[ V^{\rm C}_{ij} + V^{\rm B}_{ij} \right] - \sum_i V_{\rm scr}(r_i) \, .
\end{align}

In concluding the discussion of the interelectronic-interaction effects, we should note that the \textit{ab initio} QED calculations of the contributions arising from the diagrams in Fig.~\ref{fig:diagr} are performed using both Feynman and Coulomb gauges for the photon propagators responsible for the electron-electron interaction. For isolated states in He- and Li-like uranium, this provides an additional check of the employed numerical codes. For Be-like uranium, however, the situation is much more interesting. The eigenvalues of the matrix of the effective Hamiltonian~$H$ are nonlinearly related to the elements of the matrix itself. Different PT orders for diagonal and off-diagonal matrix elements are mixed during the diagonalization. As a result, individual contributions to the effective Hamiltonian of a certain order may vary from gauge to gauge. In Ref.~\cite{Malyshev:2021:652}, we have demonstrated how the gauge invariance of the eigenvalues is restored order by order. Keeping this in mind and also being aware of the fact that the DCB Hamiltonian is naturally related to the Coulomb gauge~\cite{Shabaev:1993:4703}, in the final compilations of the results for Be-like uranium we employ namely the Coulomb-gauge results. The Feynman-gauge calculations are used solely for control.

The higher-order screened QED contributions, which are described by the Feynman diagrams similar to the ones in the third line in Fig.~\ref{fig:diagr} but supplemented with an additional electron-electron interaction photon propagator, are inaccessible by the modern bound-state QED methods. In the present work, we estimate them approximately using the model-QED operator suggested in Ref.~\cite{Shabaev:2013:012513} and realized as the Fortran package (\texttt{QEDMOD}) in Ref.~\cite{Shabaev:2015:175:2018:69:join_pr}. For this aim, we insert the model-QED operator into the Dirac equation~(\ref{eq:DirEq}) and solve it to find the related finite-basis set. The obtained one-electron basis is then employed for the calculations of the two-photon-exchange contribution in the Breit approximation. The desired correction is separated by taking the difference with the corresponding contribution evaluated without the model-QED operator. Previously, this approach was successfully applied to the calculations of Be-like molybdenum and xenon in Refs.~\cite{Malyshev:2021:183001, Malyshev:2021:652}.

In addition, we take into account the contributions to the $2p_{3/2} \rightarrow 2s$ transitions in He-, Li-, and Be-like uranium which are beyond the Furry-picture approximation. First, we consider the nuclear recoil effect. The fully relativistic description of the nuclear recoil effect on binding energies requires application of bound-state QED, and the corresponding theory to first order in the electron-to-nucleus mass ratio~$m/M$, to zeroth order in $\alpha$, and to all orders in $\alpha Z$ was developed in Refs.~\cite{Shabaev:1985:394, Shabaev:1988:107, Shabaev:1998:59}, see also Refs.~\cite{Pachucki:1995:1854, Yelkhovsky:Budker, Adkins:2007:042508}. The lowest-order relativistic limit of this theory leads to the mass shift (MS) Hamiltonian~\cite{Shabaev:1985:394, Shabaev:1988:107, Palmer:1987:5987},
\begin{align}
\label{eq:H_MS}
H_{\rm MS} = \frac{1}{2M} \sum_{i,j} 
\left\{ 
\bp_i \cdot \bp_j
-\frac{\alpha Z}{r_i} 
\left[ 
\balpha_i + \frac{(\balpha_i\cdot\br_i)\br_i}{r_i^2}
\right] \cdot \bp_j
\right\} \, .
\end{align}
We treat the MS operator by means of PT, evaluating the first-order interelectronic-interaction correction within the Breit approximation. The calculations started with the different potentials~$V$ in Eq.~(\ref{eq:DirEq}) demonstrate the convergence of the results at the required level of accuracy. The nuclear recoil effect beyond the MS operator, which we refer to as the QED nuclear recoil effect, is taken into account within the independent-electron approximation. We note that the formalism of Refs.~\cite{Shabaev:1985:394, Shabaev:1988:107, Shabaev:1998:59} was derived initially for the point-nucleus case. In Ref.~\cite{Shabaev:1998:4235}, it was argued that the dominant part of the corresponding FNS correction can be taken into account by using the potential of the extended nucleus in the calculations, and this has been confirmed recently in Ref.~\cite{Pachucki:2023:053002}, where the most elaborated QED evaluation of the FNS correction to the nuclear recoil effect was carried out. In this work, we follow the recipe of Ref.~\cite{Shabaev:1998:4235}. Moreover, we evaluate also an additional FNS correction according to Ref.~\cite{Aleksandrov:2015:144004}. This correction differs from the one obtained in Ref.~\cite{Pachucki:2023:053002}, see the related discussion in Ref.~\cite{Anisimova:2022:062823}. Nevertheless, it turns out that for the present study these additional corrections are actually irrelevant, contributing about 1~meV to the transition energies and vanishing in the differences between the different charge states of uranium, see Sec.~\ref{sec:2}. The radiative ($\sim\!\alpha$) and second-order (in $m/M$) nuclear recoil corrections, currently available only within the $\alpha Z$-expansion approaches~\cite{Tiesinga:2021:025010, Sapirstein:1990:560, Pachucki:1995:L221}, are negligible as well. The value of the ratio~$M/m=433849$ derived from the AME2012 compilation~\cite{Wang:2012:1603} in Ref.~\cite{Yerokhin:2015:033103} is employed in the present calculations. 

Finally, we consider the nuclear polarization correction. This contribution arises from the electron-nucleus two-photon-exchange diagrams involving the excited intermediate nuclear states. We take the result from Refs.~\cite{Plunien:1995:1119:1996:4614:join_pr, Nefiodov:1996:227} and ascribe a numerical uncertainty of 50\% to it, see also Ref.~\cite{Volotka:2014:023002}.


\section{Numerical results and discussions \label{sec:2}}

\begin{table*}[t]
\centering

\renewcommand{\arraystretch}{1.25}

\caption{\label{tab:He:details} 
         Individual contributions to the $1s2p \, ^{3\!}P_{2} \rightarrow 1s2s \, ^{3\!}S_{1}$ transition energy in He-like uranium (in eV).
         See the text for details.
         }
         
\begin{tabular}{
                l@{\quad}
                l
                S[table-format=5.4]
                S[table-format=5.4]
                S[table-format=5.4]
                S[table-format=5.4]
                S[table-format=5.4]
               }
               
\hline
\hline

   \multicolumn{2}{c}{\rule{0pt}{1.2em}Contribution}    &
   \multicolumn{1}{c}{Coul}                             &
   \multicolumn{1}{c}{CH1}                              &
   \multicolumn{1}{c}{CH2}                              &
   \multicolumn{1}{c}{LDF}                              &
   \multicolumn{1}{c}{KS}                               \\ 
        
\hline   
                       
 \rule{0pt}{2.6ex}                        $E_{\rm DCB}$ &                                     & 4557.389 & 4557.382 & 4557.373 & 4557.374 & 4557.375   \\ 
                                                        &                     $E_{\rm Dirac}$ & 4527.931 & 4493.712 & 4459.613 & 4493.696 & 4524.205   \\ 
                                                        &           $E^{\rm Breit}_{\rm 1ph}$ &   30.179 &   63.718 &   97.844 &   64.115 &   34.103   \\ 
                                                        &           $E^{\rm Breit}_{\rm 2ph}$ &   -0.716 &   -0.047 &   -0.089 &   -0.444 &   -0.968   \\ 
                                                        & $E^{\rm Breit}_{\rm \geqslant 3ph}$ &   -0.005 &    0.000 &    0.005 &    0.008 &    0.035   \\ 
                                 $E^{\omega}_{\rm 1ph}$ &                                     &   -7.421 &   -7.136 &   -6.859 &   -6.878 &   -6.810   \\ 
                                           $E_{\rm np}$ &                                     &    0.039 &    0.039 &    0.039 &    0.039 &    0.039   \\ 
                              $E^{\rm Breit}_{\rm rec}$ &                                     &   -0.037 &   -0.037 &   -0.037 &   -0.037 &   -0.037   \\ 
                                          $E_{\rm QED}$ &                                     &  -40.092 &  -40.368 &  -40.636 &  -40.621 &  -40.696   \\ 
                                                        &        $E^{\rm QED}_{\text{1e-1l}}$ &  -40.991 &  -40.309 &  -39.630 &  -39.684 &  -39.341   \\ 
                                                        &        $E^{\rm QED}_{\text{1e-2l}}$ &    0.230 &    0.225 &    0.220 &    0.221 &    0.219   \\ 
                                                        &        $E^{\rm QED}_{\text{2e-1l}}$ &    0.506 &   -0.175 &   -0.852 &   -0.796 &   -1.145   \\ 
                                                        &             $E^{\rm QED}_{\rm 2ph}$ &    0.215 &   -0.065 &   -0.329 &   -0.312 &   -0.381   \\ 
                                                        &             $E^{\rm QED}_{\rm rec}$ &   -0.050 &   -0.049 &   -0.048 &   -0.048 &   -0.048   \\ 
                                                        &            $E^{\rm QED}_{\rm h.o.}$ &   -0.002 &    0.005 &    0.002 &   -0.002 &    0.000   \\ 
                                          $E_{\rm tot}$ &                                     & 4509.878 & 4509.880 & 4509.880 & 4509.877 & 4509.871   \\ 

\hline
\hline

\end{tabular}%

\end{table*}

\begin{table*}[t]
\centering

\renewcommand{\arraystretch}{1.25}

\caption{\label{tab:Li:details} 
         Individual contributions to the $1s^2 2p \, ^{2\!}P_{3/2} \rightarrow 1s^2 2s \, ^{2\!}S_{1/2}$ transition energy in Li-like uranium (in eV).
         See the text for details.
         }
         
\begin{tabular}{
                l@{\quad}
                l
                S[table-format=5.4]
                S[table-format=5.4]
                S[table-format=5.4]
                S[table-format=5.4]
                S[table-format=5.4]
               }
               
\hline
\hline

   \multicolumn{2}{c}{\rule{0pt}{1.2em}Contribution}    &
   \multicolumn{1}{c}{Coul}                             &
   \multicolumn{1}{c}{CH1}                              &
   \multicolumn{1}{c}{CH2}                              &
   \multicolumn{1}{c}{LDF}                              &
   \multicolumn{1}{c}{KS}                               \\ 
        
\hline   
                       
 \rule{0pt}{2.6ex}                        $E_{\rm DCB}$ &                                     & 4505.867 & 4505.853 & 4505.836 & 4505.839 & 4505.841   \\ 
                                                        &                     $E_{\rm Dirac}$ & 4527.931 & 4493.712 & 4459.613 & 4493.696 & 4524.205   \\ 
                                                        &           $E^{\rm Breit}_{\rm 1ph}$ &  -21.011 &   12.369 &   46.322 &   12.007 &  -18.321   \\ 
                                                        &           $E^{\rm Breit}_{\rm 2ph}$ &   -1.023 &   -0.217 &   -0.101 &    0.138 &   -0.047   \\ 
                                                        & $E^{\rm Breit}_{\rm \geqslant 3ph}$ &   -0.030 &   -0.012 &    0.002 &   -0.001 &    0.003   \\ 
                                 $E^{\omega}_{\rm 1ph}$ &                                     &   -7.401 &   -7.123 &   -6.852 &   -6.870 &   -6.803   \\ 
                                           $E_{\rm np}$ &                                     &    0.039 &    0.039 &    0.039 &    0.039 &    0.039   \\ 
                              $E^{\rm Breit}_{\rm rec}$ &                                     &   -0.036 &   -0.036 &   -0.036 &   -0.036 &   -0.036   \\ 
                                          $E_{\rm QED}$ &                                     &  -38.912 &  -39.169 &  -39.419 &  -39.405 &  -39.475   \\ 
                                                        &        $E^{\rm QED}_{\text{1e-1l}}$ &  -40.991 &  -40.309 &  -39.630 &  -39.684 &  -39.341   \\ 
                                                        &        $E^{\rm QED}_{\text{1e-2l}}$ &    0.230 &    0.225 &    0.220 &    0.221 &    0.219   \\ 
                                                        &        $E^{\rm QED}_{\text{2e-1l}}$ &    1.642 &    0.953 &    0.268 &    0.324 &   -0.023   \\ 
                                                        &             $E^{\rm QED}_{\rm 2ph}$ &    0.294 &    0.027 &   -0.226 &   -0.211 &   -0.278   \\ 
                                                        &             $E^{\rm QED}_{\rm rec}$ &   -0.050 &   -0.049 &   -0.048 &   -0.048 &   -0.048   \\ 
                                                        &            $E^{\rm QED}_{\rm h.o.}$ &   -0.037 &   -0.016 &   -0.003 &   -0.007 &   -0.004   \\ 
                                          $E_{\rm tot}$ &                                     & 4459.558 & 4459.564 & 4459.568 & 4459.567 & 4459.565   \\ 

\hline
\hline

\end{tabular}%

\end{table*}

\begin{table*}[t]
\centering

\renewcommand{\arraystretch}{1.25}

\caption{\label{tab:Be:details} 
         Gradual change of the $1s^2 2s2p \, ^{1\!}P_{1} \rightarrow 1s^2 2s^2 \, ^{1\!}S_{0}$ transition energy in Be-like uranium (in eV) as various contributions are successively taken into account in the effective Hamiltonians~$H$ for the initial and final states.
         The corresponding increments are shown in italics under the values.
         The individual contributions are not exactly additive in this case.
         See the text for details.
         }          
         
\begin{tabular}{
                l
                r
                r
                r
                r
                r
               }
               
\hline
\hline

   \multicolumn{1}{c}{\rule{0pt}{1.2em}Contribution~~}    &
   \multicolumn{1}{c}{\parbox{1.6cm}{~~~Coul}}            &
   \multicolumn{1}{c}{\parbox{1.6cm}{~~CH1}}              &
   \multicolumn{1}{c}{\parbox{1.6cm}{~~CH2}}              &
   \multicolumn{1}{c}{\parbox{1.6cm}{~~LDF}}              &
   \multicolumn{1}{c}{\parbox{1.6cm}{~~~KS}}              \\  
        
\hline   
                       
 \rule{0pt}{2.6ex}                   $E_{\rm DCB}$ &         4546.778 &         4546.759 &         4546.739 &         4546.743 &         4546.744   \\ 
                           $+H_{\rm 1ph}^{\omega}$ &         4539.392 &         4539.652 &         4539.903 &         4539.888 &         4539.956   \\ 
                                                   & $ \it   -7.385 $ & $ \it   -7.107 $ & $ \it   -6.836 $ & $ \it   -6.855 $ & $ \it   -6.788 $   \\ 
                                     $+H_{\rm np}$ &         4539.431 &         4539.690 &         4539.941 &         4539.927 &         4539.994   \\ 
                                                   & $ \it    0.038 $ & $ \it    0.038 $ & $ \it    0.038 $ & $ \it    0.038 $ & $ \it    0.038 $   \\ 
                        $+H^{\rm Breit}_{\rm rec}$ &         4539.396 &         4539.655 &         4539.906 &         4539.892 &         4539.959   \\ 
                                                   & $ \it   -0.035 $ & $ \it   -0.035 $ & $ \it   -0.035 $ & $ \it   -0.035 $ & $ \it   -0.035 $   \\ 
                     $+H^{\rm QED}_{\text{1e-1l}}$ &         4499.522 &         4500.444 &         4501.354 &         4501.287 &         4501.686   \\ 
                                                   & $ \it  -39.874 $ & $ \it  -39.211 $ & $ \it  -38.552 $ & $ \it  -38.605 $ & $ \it  -38.274 $   \\ 
                     $+H^{\rm QED}_{\text{1e-2l}}$ &         4499.745 &         4500.662 &         4501.568 &         4501.501 &         4501.898   \\ 
                                                   & $ \it    0.223 $ & $ \it    0.218 $ & $ \it    0.214 $ & $ \it    0.214 $ & $ \it    0.212 $   \\ 
                     $+H^{\rm QED}_{\text{2e-1l}}$ &         4501.513 &         4501.761 &         4502.000 &         4501.989 &         4502.053   \\ 
                                                   & $ \it    1.768 $ & $ \it    1.098 $ & $ \it    0.433 $ & $ \it    0.488 $ & $ \it    0.155 $   \\ 
                          $+H^{\rm QED}_{\rm 2ph}$ &         4501.851 &         4501.835 &         4501.825 &         4501.828 &         4501.826   \\ 
                                                   & $ \it    0.337 $ & $ \it    0.074 $ & $ \it   -0.176 $ & $ \it   -0.161 $ & $ \it   -0.227 $   \\ 
                          $+H^{\rm QED}_{\rm rec}$ &         4501.802 &         4501.787 &         4501.778 &         4501.781 &         4501.780   \\ 
                                                   & $ \it   -0.049 $ & $ \it   -0.048 $ & $ \it   -0.047 $ & $ \it   -0.047 $ & $ \it   -0.046 $   \\ 
                         $+H^{\rm QED}_{\rm h.o.}$ &         4501.760 &         4501.768 &         4501.773 &         4501.772 &         4501.772   \\ 
                                                   & $ \it   -0.041 $ & $ \it   -0.019 $ & $ \it   -0.006 $ & $ \it   -0.009 $ & $ \it   -0.008 $   \\ 

\hline
\hline

\end{tabular}%

\end{table*}

\begin{table*}[t]
\centering

\renewcommand{\arraystretch}{1.25}

\caption{\label{tab:error} 
         Error budget for the transition energies and their differences (in eV).
         See the text for details.
         }
         
\begin{tabular}{
                l
                S[table-format=1.3]                
                S[table-format=1.3] 
                S[table-format=1.3] 
                S[table-format=1.3] 
                S[table-format=1.3] 
                S[table-format=1.3]                      
               }
               
\hline
\hline

   \multicolumn{1}{c}{\rule{0pt}{1.2em}\parbox{2.9cm}{\centering Source}}          &
   \multicolumn{1}{c}{\parbox{1.4cm}{\centering He}}                               &
   \multicolumn{1}{c}{\parbox{1.4cm}{\centering Li}}                               &
   \multicolumn{1}{c}{\parbox{1.4cm}{\centering Be}}                               &
   \multicolumn{1}{c}{\parbox{1.4cm}{\centering $\text{He}-\text{Li}$}}            &
   \multicolumn{1}{c}{\parbox{1.4cm}{\centering $\text{He}-\text{Be}$}}            &
   \multicolumn{1}{c}{\parbox{1.4cm}{\centering $\text{Be}-\text{Li}$}}           \\
        
\hline   

1el-FNS                   &  0.034  &  0.034  &  0.034  &  {---}  &  0.003  &  0.003  \\

nucl. pol.                &  0.020  &  0.020  &  0.020  &  {---}  &  0.002  &  0.002  \\
  
1el-2loop                 &  0.086  &  0.086  &  0.086  &  {---}  &  0.009  &  0.009  \\
  
h.o. QED, 2loop           &  0.030  &  0.030  &  0.030  &  0.030  &  0.030  &  0.030  \\

h.o. QED, $\geqslant$~3ph &  0.017  &  0.018  &  0.021  &  0.018  &  0.021  &  0.021  \\

h.o. QED, scatter         &  0.002  &  0.016  &  0.021  &  0.014  &  0.019  &  0.005  \\  

Other                     &  0.008  &  0.008  &  0.009  &  0.008  &  0.008  &  0.007  \\

Total                     &  0.10   &  0.10   &  0.10   &  0.039  &  0.043  &  0.039  \\

\hline 

\end{tabular}%

{
\begin{flushleft}

\end{flushleft}
}

\end{table*}

In this section, we present our theoretical predictions for the $2p_{3/2} \rightarrow 2s$ transition energies in He-, Li-, and Be-like uranium as well as for all possible their differences. First, we discuss individual contributions to the transition energies and demonstrate the convergence of the QED perturbation series. Second, we give an error budget of the calculations and estimate the uncertainties related to the uncalculated higher-order contributions. Finally, we compare the obtained values with the results of previous calculations and available experimental data.

The individual contributions to the $1s2p \, ^{3\!}P_{2} \rightarrow 1s2s \, ^{3\!}S_{1}$ transition in He-like uranium and $1s^2 2p \, ^{2\!}P_{3/2} \rightarrow 1s^2 2s \, ^{2\!}S_{1/2}$ transition in Li-like uranium are presented in Tables~\ref{tab:He:details} and \ref{tab:Li:details}, respectively. As noted in Sec.~\ref{sec:1}, standard single-level PT is used for these ions. In Tables~\ref{tab:He:details} and \ref{tab:Li:details}, the first line labeled~$E_{\rm DCB}$ shows the values obtained in the Breit approximation by means of the CI method. The subsequent rows, $E_{\rm Dirac}$, $E^{\rm Breit}_{\rm 1ph}$, $E^{\rm Breit}_{\rm 2ph}$, and $E^{\rm Breit}_{\rm \geqslant 3ph}$, present the contributions to $E_{\rm DCB}$ of zeroth, first, second, and third and higher orders with respect to the interaction operator~$V_{\rm int}$ in Eq.~(\ref{eq:DCB}), correspondingly. The zeroth-order term~$E_{\rm Dirac}$ includes the nuclear deformation correction. We draw attention to the fact that the total values~$E_{\rm DCB}$ vary slightly from potential to potential. These variations are not due to numerical errors, but reflect a dependence on the choice of the projectors~$\Lambda^{(+)}$ in Eq.~(\ref{eq:DCB}). The ambiguity of the Breit-approximation results can be eliminated only in the framework of the rigorous QED approach, see the discussion in Ref.~\cite{Kozhedub:2019:062506}. The next line~$E_{\rm 1ph}^{\omega}$ is the frequency-dependent correction to the Breit interaction. It is obtained as the difference between the contribution of the one-photon-exchange diagram, rigorously evaluated within the QED approach, and the corresponding Breit-approximation result. This correction is often included into the DCB Hamiltonian. We do not follow this approximate recipe and, instead, consider the contributions of the one- and two-photon-exchange diagrams within the bound-state QED approach. The line labeled~$E_{\rm np}$ gives the nuclear polarization correction. The next row~$E^{\rm Breit}_{\rm rec}$ shows the part of the nuclear recoil effect evaluated with the use of the MS operator taking into account the additional FNS correction from Ref.~\cite{Aleksandrov:2015:144004}. In what follows, we refer to the sum of the listed above terms as the non-QED contribution. The line~$E_{\rm QED}$ presents the total QED contribution. First, it includes the one-electron one-loop contribution (the second and the third diagrams in the first row in Fig.~\ref{fig:diagr}), $E^{\rm QED}_{\text{1e-1l}}$, the one-electron two-loop contribution, $E^{\rm QED}_{\text{1e-2l}}$, the two-electron one-loop contribution (the third row in Fig.~\ref{fig:diagr}), $E^{\rm QED}_{\text{2e-1l}}$, and the part of the two-photon-exchange contribution beyond the Breit approximation, $E^{\rm QED}_{\rm 2ph}$ (the second row in Fig.~\ref{fig:diagr}). Second, $E_{\rm QED}$ comprises the QED part of the nuclear recoil contribution, $E^{\rm QED}_{\rm rec}$. We note that for the $2p_{3/2} \rightarrow 2s$ transition in highly charged uranium the QED term~$E^{\rm QED}_{\rm rec}$ even exceeds the corresponding Breit-approximation value, $E^{\rm Breit}_{\rm rec}$. If necessary, the QED contribution to the nuclear recoil effect can also be evaluated beyond the independent-electron approximation with the help of the formalism derived recently in Refs.~\cite{Malyshev:2019:012510, Malyshev:2020:052506}. However, since the dominant uncertainties arise nowadays from other sources, see the discussion below, we have not done this yet. Third, the total QED contribution involves also the higher-order screened QED correction~$E^{\rm QED}_{\rm h.o.}$ approximately treated by means of the model-QED operator. In the last row, the total transition energies are given. From Tables~\ref{tab:He:details} and \ref{tab:Li:details}, it is seen that, although the individual contributions may vary significantly from column to column, the total results are almost independent on the initial approximation. In Table~\ref{tab:He:details}, the KS potential leads to the value which is slightly away from the others. Apparently, the reason is that this screening potential, behaving like $(3\alpha)/r$ at the large distances, is poorly suited for He-like uranium. Moreover, analyzing Table~\ref{tab:He:details} as a whole, one can conclude that the application of the extended Furry picture in the case of He-like uranium does not appear to provide benefit compared to the calculations with the Coulomb potential of the nucleus. Nevertheless, the benefits become evident already for Li-like uranium. In Table~\ref{tab:Li:details}, the total values obtained for the screening potentials are shifted slightly with respect to the Coulomb ones, which is a result of the PT rearrangement. We note also that this shift increases when the contribution~$E^{\rm QED}_{\rm h.o.}$ is omitted.

The individual contributions to the $1s^2 2s2p \, ^{1\!}P_{1} \rightarrow 1s^2 2s^2 \, ^{1\!}S_{0}$ transition in Be-like uranium are presented in Table~\ref{tab:Be:details}. In contrast to He- and Li-like ions, for Be-like uranium we employ PT for quasidegenerate levels. The binding energies of the initial and final states are obtained as the eigenvalues of the matrices for the corresponding effective Hamiltonians. Strictly speaking, the diagonalization procedure makes the different terms not quite additive. For this reason, in Table~\ref{tab:Be:details} we demonstrate how the value of the transition energy changes as we successively take into account different effects. The notations are similar to the ones used in Tables~\ref{tab:He:details} and \ref{tab:Li:details}. For instance, the first row~$E_{\rm DCB}$ gives the results of the Breit-approximation calculations by means of the CI method, while the line labeled~$+H_{\rm 1ph}^{\omega}$ presents the values obtained after the frequency-dependent correction to the one-photon-exchange contribution has been additionally included into the effective Hamiltonians. We note that the matrix $H_{\rm CI}$, obtained in accordance with the method of Refs.~\cite{Malyshev:2019:010501_R, Kozhedub:2019:062506, Malyshev:2021:652} from the CI calculations for a set of quasidegenerate levels, naturally has the eigenvalues coinciding with the corresponding CI energies. However, the zeroth-order contribution to $H_{\rm CI}$ has been modified so to include the nuclear deformation correction. Therefore, the values in the line~$E_{\rm DCB}$, obtained by diagonalizing $H_{\rm CI}$, take into account this effect. For convenience, the successive increments of the results are shown in Table~\ref{tab:Be:details} in italics under the corresponding values. The total transition energies are given in the row labeled~$+H^{\rm QED}_{\rm h.o.}$, which sums up all the contributions. As in the case of He- and Li-like uranium, the total energies obtained for the different potentials agree well with each other. It is worth noting that the one-electron contributions for He- and Li-like ions, e.g., the one-electron one-loop contributions, coincide for these charge states but differ from the corresponding terms for Be-like uranium. This is the direct manifestation of the level mixing, which is taken into account by means of PT for quasidegenerate levels. Finally, returning to the question of the dependence of the Breit approximation on the choice of the operators~$\Lambda^{(+)}$ in Eq.~(\ref{eq:DCB}), we note that the maximum absolute values of the deviations between the results given in the first lines in Tables~\ref{tab:He:details}, \ref{tab:Li:details}, and \ref{tab:Be:details} constitute $16$, $31$, and $39$~meV, respectively. The corresponding quantities evaluated for the last rows are equal to $9$, $11$, and $12$~meV for He-, Li-, and Be-like uranium, correspondingly. This demonstrates how the QED approach eliminates the ambiguity related to $\Lambda^{(+)}$.

An essential part of the present work is the careful estimation of the uncertainties of the obtained theoretical predictions. Table~\ref{tab:error} presents the error budget of our calculations. The columns labeled ``He'', ``Li'', and ``Be'' refer to the transition energies in the corresponding charge states of uranium, while the remaining columns concern their differences. In Table~\ref{tab:error}, the first line~``1el-FNS'' shows the uncertainties due to the FNS effect on the one-electron Dirac energies. They are obtained from the calculations of the nuclear deformation correction. In the ``$\text{He}-\text{Li}$'' difference, this uncertainty vanishes, since the corresponding contributions to the transitions in He- and Li-like uranium completely cancel each other. For the transition-energy differences involving Be-like uranium, this is not the case because of the level mixing. For the ground state, we find the admixture of the configuration~$1s^2 2p_{1/2}^2$ to the dominant one, $1s^2 2s^2$, at the level of few percents. When evaluating the ``1el-FNS'' uncertainties for the ``$\text{He}-\text{Be}$'' and ``$\text{Be}-\text{Li}$'' differences, we have conservatively estimated the uncompensated remainders of the one-electron contributions to be of the order of 10\%. This leads to considerably reduced but nonetheless nonzero uncertainties indicated in Table~\ref{tab:error}. The same situation holds for all other one-electron sources of the uncertainties. The second line~``nucl. pol.'' in Table~\ref{tab:error} gives the uncertainties related to the nuclear polarization effect. As noted in Sec.~\ref{sec:1}, these contributions are estimated within the independent-electron approximation, and they are ascribed the uncertainty of 50\%. The third row~``1el-2loop'' shows the uncertainties of the one-electron two-loop contributions taken from Ref.~\cite{Yerokhin:2015:033103}. The large values of the latter uncertainties are related with the fact that the nonperturbative (in $\alpha Z$) calculations are not fully completed yet for some one-electron two-loop diagrams. Namely, the self-energy diagram with a fermion loop inserted into the photon propagator and the vacuum-polarization diagram with an additional photon line within the fermion loop are evaluated using the free-loop approximation. The uncertainty due to this approximation determines currently the largest theoretical error for the Lamb shift in H-like uranium as well. The next three lines in Table~\ref{tab:error} present our various estimations of the uncalculated higher-order QED contributions. First, the screening of the one-electron two-loop contribution is estimated by multiplying the corresponding term for the $1s$ state by a conservative factor of $2/Z$. This uncertainty is ascribed for both transition energies and their differences, and it is indicated in the line with the label ``h.o. QED, 2loop''. Second, the QED correction to the interelectronic-interaction contribution due to the exchange by three or more photons is obtained according to the procedure from Ref.~\cite{Kozhedub:2019:062506}. For the transition energies, the absolute value of this uncertainty is conservatively estimated as the term $E^{\rm Breit}_{\rm \geqslant 3ph}$ multiplied by a factor of $2 E^{\rm QED}_{\rm 2ph} / E^{\rm Breit}_{\rm 2ph}$. In order to avoid an underestimation due to the abnormal smallness of $E^{\rm Breit}_{\rm \geqslant 3ph}$, we evaluate these uncertainties using the Coulomb-potential results for the ground states corresponding to the ion under consideration, i.e., we do this for the $1s^2 \, ^{1\!}S_{0}$, $1s^2 2s \, ^{2\!}S_{1/2}$, and $1s^2 2s^2 \, ^{1\!}S_{0}$ states in the case of He-, Li-, and Be-ions, respectively. For the differences of the transition energies, we take the maximum uncertainty of the charge states involved. These uncertainties are shown in the row~``h.o. QED, $\geqslant$~3ph''. Third, we take into account the scatter of the results obtained. Namely, for all the considered quantities, we sum up the absolute value of the $E_{\rm QED}^{\rm h.o.}$ contribution for the LDF potential and the absolute value of the difference between the total results for the LDF and Coulomb potentials. The values obtained in this way are given in the line~``h.o. QED, scatter''. The row labeled~``Other'' represents all other sources of the uncertainties. In particular, it includes the numerical errors of the calculations and the uncertainty due to the remaining part of the FNS effect. The numerical errors are estimated by studying the convergence of the results with respect to the partial-wave expansions and the sizes of the finite-basis sets used. We stress that, in contrast to the one-electron two-loop diagrams, the one- and two-electron one-loop contributions as well as the contributions of the two-photon-exchange diagrams are evaluated to all orders in $\alpha Z$, and their uncertainties are purely numerical and practically negligible. As for the CI calculations, their convergence has been analyzed employing the approach described in Ref.~\cite{Kaygorodov:2019:032505}. The uncertainty of the contribution $E^{\rm Breit}_{\rm \geqslant 3ph}$ is found to be negligible. Apart from that, the line~``Other'' comprises two following errors. First, the QED nuclear recoil effect beyond the independent-electron approximation is conservatively estimated by multiplying the leading QED recoil contribution for the $1s$ state by a factor of $2/Z$. The resulting uncertainty constitutes about 5~meV, and it is ascribed for the transition energies as well as their differences. Second, we have applied a similar procedure for the nuclear polarization effect, which results in the uncertainty of the order of 4~meV. Obviously, treating this uncertainty for the transition energies against the background of 20~meV shown in the second line of Table~\ref{tab:error} is meaningless. Nevertheless, we consider it important to take into account this error for the transition-energy differences. The screened nuclear polarization correction was rigorously studied in Ref.~\cite{Volotka:2014:023002}, and for the $1s^2 2p \, ^{2\!}P_{3/2} \rightarrow 1s^2 2s \, ^{2\!}S_{1/2}$ transition in U$^{89+}$ it constitutes $-1.9$~meV which confirms the reliability of our estimation. The total uncertainties are obtained by summing quadratically all the discussed errors. From Table~\ref{tab:error}, it can be seen that the transition energies in all considered charge states of uranium possess the theoretical uncertainties of 0.10~eV. Currently, the accuracy of the calculations is determined by the one-electron two-loop contribution. For the differences of the transition energies, this uncertainty is reduced, and the higher-order QED effects not covered so far by the state-of-the-art bound-state QED methods come to the fore.

Let us proceed with the discussion of the obtained theoretical predictions. The results for the LDF potential are chosen as the final ones for all the charge states of uranium. In principle, the specific choice of the screening potential does not make a big difference in view of a good convergence of the PT series demonstrated in Tables~\ref{tab:He:details}, \ref{tab:Li:details}, and \ref{tab:Be:details}. The uncertainty~``h.o. QED, scatter'' in Table~\ref{tab:error} covers the scatter of the total results. Moreover, we note that the means of the values obtained for the different potentials almost coincide with the LDF results, even if one takes into account the calculations for the Coulomb potential.

\begin{table}[t]
\centering

\renewcommand{\arraystretch}{1.25}

\caption{\label{tab:He:comparison} 
         The $1s2p \, ^{3\!}P_{2} \rightarrow 1s2s \, ^{3\!}S_{1}$ transition energy in He-like uranium (in eV).
         }
         
\begin{tabular}{
                S[table-format=4.2(2), table-space-text-post=$^{\,a}$, table-align-text-post=false]
                c                
                c              
                l
               }
               
\hline
\hline

   \multicolumn{1}{c}{\rule{0pt}{1.2em}Energy}    &
   \multicolumn{1}{c}{Th./Expt.}                  &
   \multicolumn{1}{c}{~~Year~~}                   &
   \multicolumn{1}{c}{Reference}                  \\      
        
\hline   

 4509.88(10)$^a$    &  Th.    &  2023  &  This work    \\
 
 4509.71(99)        &  Expt.  &  2009  &  Trassinelli \textit{et al.} \cite{Trassinelli:2009:63001} \\
 
 4509.97(26)$^b$    &  Th.    &  2005  &  Artemyev \textit{et al.} \cite{Artemyev:2005:062104} \\
 
 4510.05(24)$^c$    &  Expt.  &  1996  &  Beiersdorfer \textit{et al.} \cite{Beiersdorfer:1996:4000} \\
 
 4510.46            &  Th.    &  1994  &  Plante \textit{et al.} \cite{Plante:1994:3519} \\
 
 4510.65            &  Th.    &  1993  &  Chen \textit{et al.} \cite{Chen:1993:3692} \\
 
 4510.0$(1.0)$      &  Th.    &  1988  &  Drake \cite{Drake:1988:586} \\

\hline 

\end{tabular}%

{
\begin{flushleft}
The results of Refs.~\cite{Drake:1988:586, Chen:1993:3692, Plante:1994:3519} do not include a correction due to the nuclear polarization effect. \\
$^{a}$ In Ref.~\cite{Kozhedub:2019:062506}, the value of $4509.88(11)$ was obtained. \\
$^{b}$ The result is corrected for the updated root-mean-square nuclear radius~\cite{Angeli:2013:69}. \\
$^{c}$ The transition was not identified unambiguously, and a possible blending with the transition in U$^{88+}$ was discussed. 
\end{flushleft}
}

\end{table}

\begin{table}[t]
\centering

\renewcommand{\arraystretch}{1.25}

\caption{\label{tab:Li:comparison} 
         The $1s^2 2p \, ^{2\!}P_{3/2} \rightarrow 1s^2 2s \, ^{2\!}S_{1/2}$ transition energy in Li-like uranium (in eV).
         }
         
\begin{tabular}{
                S[table-format=4.2(2), table-align-text-post=false, table-space-text-post=$^{\,a}$]
                c                
                c              
                l
               }
               
\hline
\hline

   \multicolumn{1}{c}{\rule{0pt}{1.2em}Energy}    &
   \multicolumn{1}{c}{Th./Expt.}                  &
   \multicolumn{1}{c}{~~Year~~}                   &
   \multicolumn{1}{c}{Reference}                  \\      
        
\hline   

 4459.57(10)    &  Th.    &  2023  &  This work    \\
 
 4459.46(8)     &  Th.    &  2011  &  Sapirstein and Cheng \cite{Sapirstein:2011:012504}  \\
 
 4459.48        &  Th.    &  2000  &  Cheng \textit{et al.} \cite{Cheng:2000:054501} \\
 
 4460.00        &  Th.    &  1998  &  Santos \textit{et al.} \cite{Santos:1998:149} \\
 
 4459.55$^a$    &  Expt.  &  1996  &  Beiersdorfer \textit{et al.} \cite{Beiersdorfer:1996:4000} \\
 
 4459.14$^b$    &  Th.    &  1995  &  Chen \textit{et al.} \cite{Chen:1995:266} \\
 
 4459.49$^c$    &  \rdelim\}{2}{14.7mm}[\,\,\,\,Th.]        & 
                   \multirow{2}{*}{1995}                    & 
                   \multirow{2}{*}{Johnson \textit{et al.} \cite{Johnson:1995:297}} \\
 
 4459.13$^c$    &  \\
 
 4459.37(21)    &  Expt.  &  1993  &  Beiersdorfer \textit{et al.} \cite{Beiersdorfer:1993:3939, Beiersdorfer:1995:114} \\
 
 4459.10        &  Th.    &  1991  &  Kim \textit{et al.} \cite{Kim:1991:148} \\
 
 4459.94        &  Th.    &  1979  &  Cheng \textit{et al.} \cite{Cheng:1979:111} \\

\hline 

\end{tabular}%

{
\begin{flushleft}
The results of Refs.~\cite{Cheng:1979:111, Kim:1991:148, Johnson:1995:297, Cheng:2000:054501} do not include a correction due to the nuclear polarization effect. \\
$^a$ The value was found in the experiment aimed at the measurement of the $1s2p \, ^{3\!}P_{2} \rightarrow 1s2s \, ^{3\!}S_{1}$ transition in He-like uranium. The uncertainty was not specified. \\
$^b$ The nuclear polarization correction is corrected from 0.21~eV to 0.04~eV \cite{Plunien:1995:1119:1996:4614:join_pr, Nefiodov:1996:227}. \\
$^c$ The calculations were performed for two different model potentials.
\end{flushleft}
}

\end{table}

\begin{table}[t]
\centering

\renewcommand{\arraystretch}{1.25}

\caption{\label{tab:Be:comparison} 
         The $1s^2 2s2p \, ^{1\!}P_{1} \rightarrow 1s^2 2s^2 \, ^{1\!}S_{0}$ transition energy in Be-like uranium (in eV).
         }
         
\begin{tabular}{
                S[table-format=4.2(2), table-align-text-post=false, table-space-text-post=$^{\,a}$]
                c                
                c              
                l
               }
               
\hline
\hline

   \multicolumn{1}{c}{\rule{0pt}{1.2em}Energy}    &
   \multicolumn{1}{c}{Th./Expt.}                  &
   \multicolumn{1}{c}{~~Year~~}                   &
   \multicolumn{1}{c}{Reference}                  \\      
        
\hline   

 4501.77(10)    &  Th.    &  2023  &  This work    \\
 
 4502.8$(12.8)$ &  Th.    &  2019  &  Kaygorodov \textit{et al.} \cite{Kaygorodov:2019:032505} \\
 
 4501.66        &  Th.    &  2008  &  Cheng \textit{et al.} \cite{Cheng:2008:052504} \\
 
 4501.73        &  Th.    &  2000  &  Cheng \textit{et al.} \cite{Cheng:2000:054501} \\
 
 4501.54        &  Th.    &  1998  &  Santos \textit{et al.} \cite{Santos:1998:149} \\
 
 4501.36        &  Th.    &  1997  &  Chen and Cheng \cite{Chen:1997:166} \\
 
 4501.602       &  Th.    &  1996  &  Safronova \textit{et al.} \cite{Safronova:1996:4036} \\
 
 4502.88$^a$    &  \rdelim\}{2}{14.7mm}[\,\,\,\,Th.$^b$]    & 
                   \multirow{2}{*}{1995}                    & 
                   \multirow{2}{*}{Johnson \textit{et al.} \cite{Johnson:1995:297}} \\
 
 4504.28$^a$    &  \\ 
 
 4501.72(27)    &  Expt.  &  1993  &  Beiersdorfer \textit{et al.} \cite{Beiersdorfer:1993:3939} \\

\hline 

\end{tabular}%

{
\begin{flushleft}
The results of Refs.~\cite{Johnson:1995:297, Safronova:1996:4036, Chen:1997:166, Cheng:2000:054501, Cheng:2008:052504, Kaygorodov:2019:032505} do not include a correction due to the nuclear polarization effect. \\
$^a$ The calculations were performed for two different model potentials. \\
$^b$ The authors indicated a poor convergence of the many-body perturbation theory for U$^{88+}$.
\end{flushleft}
}

\end{table}

We present our theoretical predictions for the $1s2p \, ^{3\!}P_{2} \rightarrow 1s2s \, ^{3\!}S_{1}$ transition in He-like uranium, $1s^2 2p \, ^{2\!}P_{3/2} \rightarrow 1s^2 2s \, ^{2\!}S_{1/2}$ transition in Li-like uranium, and $1s^2 2s2p \, ^{1\!}P_{1} \rightarrow 1s^2 2s^2 \, ^{1\!}S_{0}$ transition in Be-like uranium in Tables~\ref{tab:He:comparison}, \ref{tab:Li:comparison}, and \ref{tab:Be:comparison}, respectively. The obtained values are compared with the results of the previous calculations and available experimental data. We do not give here a detailed review on theoretical methods employed in the cited references, but instead refer the reader to the original works. We note merely that the many-electron QED effects were rigorously treated only in Refs.~\cite{Artemyev:2005:062104, Kozhedub:2019:062506} and Ref.~\cite{Sapirstein:2011:012504} for He- and Li-like uranium, correspondingly. For Be-like uranium, calculations of this kind are absent in the literature. In Ref.~\cite{Kozhedub:2019:062506}, we obtained $4509.88(11)$~eV for He-like uranium. The present calculations have led to the same value, whereas the comprehensive analysis of the error budget has resulted in the slightly reduced uncertainty of 0.10~eV. Good agreement is found with the results of Ref.~\cite{Artemyev:2005:062104} as well. It should be noted that in Ref.~\cite{Artemyev:2005:062104} the calculations were performed for the RMS radius equal to $5.8507$~fm instead of $5.8571$~fm utilized in the present work. In Table~\ref{tab:He:comparison}, the value from Ref.~\cite{Artemyev:2005:062104} has been appropriately corrected. In Ref.~\cite{Kozhedub:2019:062506}, the corresponding corrections were introduced, by mistake, with the opposite sign, which increased the discrepancy between the results. In the present work, we fix this mistake and emphasize the better agreement with Ref.~\cite{Artemyev:2005:062104}. In the case of Li-like uranium, our result~$4459.57(10)$~eV shows good agreement with the value~$4459.46(8)$~eV obtained by Sapirstein and Cheng in Ref.~\cite{Sapirstein:2011:012504}. Meanwhile, almost half of the deviation of $0.11$~eV is caused by the difference in the FNS-effect treatment. Indeed, the Fermi model with the parameter~$c$ equal to $7.137\,53$~fm was used in Ref.~\cite{Sapirstein:2011:012504} for uranium. This corresponds to the RMS radius $R=5.8610$~fm. If we reevaluate the zeroth-order contribution according to our RMS radius, the value of Ref.~\cite{Sapirstein:2011:012504} increases by about $35$~meV. The nuclear deformation correction, not taken into account there, additionally makes it higher by $20$~meV. Finally, the larger uncertainty of our result is explained by a more conservative way to estimate it. In particular, the results of Refs.~\cite{Yerokhin:2006:253004, Yerokhin:2009:040501_R} were employed in Ref.~\cite{Sapirstein:2011:012504} to take into account the one-electron two-loop contribution, and the resulting value constituted $0.212(53)$~eV. We utilize the more recent compilation~\cite{Yerokhin:2015:033103, Yerokhin:2018:052509} and retrieve from it the value~$0.230(86)$~eV (for the Coulomb potential). Replacing the error shown in the third line in Table~\ref{tab:error} with $53$~meV would reduce our total uncertainty for U$^{89+}$ from $0.10$~eV to $0.08$~eV. For Be-like ions, the $1s^2 2s2p \, ^{1\!}P_{1} \rightarrow 1s^2 2s^2 \, ^{1\!}S_{0}$ transition was studied in Ref.~\cite{Kaygorodov:2019:032505} by means of the CI method combined with the model-QED operator to approximately treat the QED effects, and the uncertainties were estimated rather conservatively. The comparison with the present calculations shows that the accuracy of the approach of Ref.~\cite{Kaygorodov:2019:032505} is at least one order of magnitude higher. Concluding the comparison of our transition energies with the results of the previous calculations and wishing to further emphasize the importance of employed approach, we would like to draw the attention to the results from Ref.~\cite{Johnson:1995:297} shown in Tables~\ref{tab:Li:comparison} and \ref{tab:Be:comparison}. In Ref.~\cite{Johnson:1995:297}, many-body perturbation theory (MBPT) up to the second order was applied to the calculations of the $2p_{3/2} \rightarrow 2s$ transition energies starting with two different local potentials. The QED corrections were treated at the one-loop level using the same potentials. For Li-like uranium, reasonable agreement takes place. However, the dependence on the initial approximation, which can be resolved only by considering the many-electron QED contributions, is clearly demonstrated. For Be-like uranium, as the authors indicate, the situation is worse. The poor convergence of MBPT was found, and the obtained results are far from the experimental data. For the electron-electron correlations, single-level PT up to the second order can readily be replaced, e.g., with the nonperturbative CI method, which solves the convergence problem. For the QED effects, the consideration of which is currently restricted exactly by the second order, it is not feasible to do so. For this reason, only QED PT for quasidegenerate levels, constructed in the present work, provides the proper \textit{ab initio} description in the case of Be-like ions. 

As for the comparison with experiment, our theoretical predictions for the transition energies are in good agreement with the available data. For He-like uranium, two measurements exist in the literature. The first one was carried out on the high-energy electron-beam ion trap (EBIT) facility at the Lawrence Livermore National Laboratory~\cite{Beiersdorfer:1996:4000}, and the second one was performed at the ESR storage ring at the GSI in Darmstadt~\cite{Trassinelli:2009:63001}. Despite the fact that for the earlier experiment on the EBIT the lower uncertainty is indicated, it has to be noted that the $1s2p \, ^{3\!}P_{2} \rightarrow 1s2s \, ^{3\!}S_{1}$ transition was not identified unambiguously in Ref.~\cite{Beiersdorfer:1996:4000}. The measurement was based on a weak signal from the transition observed with an intensity just above the level of background fluctuations. Moreover, since the EBIT simultaneously deals with the different charge states, the possibility was discussed that the $1s^2 2p^2 \, ^{3\!}P_{1} \rightarrow 1s^2 2s2p \, ^{3\!}P_{0}$ transition in Be-like uranium might blend and mask the desired $1s2p \, ^{3\!}P_{2} \rightarrow 1s2s \, ^{3\!}S_{1}$ transition. For this reason, the new storage-ring experiment, which states the uncertainty at the level of about $0.2$~eV~\cite{GSI_report_2021}, is of great interest. For Li- and Be-like uranium, the $2p_{3/2} \rightarrow 2s$ transition energies were measured on the EBIT in Refs.~\cite{Beiersdorfer:1993:3939, Beiersdorfer:1995:114}. For the $1s^2 2s2p \, ^{1\!}P_{1} \rightarrow 1s^2 2s^2 \, ^{1\!}S_{0}$ transition in Be-like uranium, our theoretical value, $4501.77(10)$~eV, and the experimental one, $4501.72(27)$~eV, are in excellent agreement with each other. We stress once more that the second-order QED contributions, rigorously evaluated in the present work, have not been considered previously for this transition in U$^{88+}$. For Li-like uranium, good agreement takes place as well. We note, however, that for U$^{89+}$ another experimental value exists. The spectra, observed in Ref.~\cite{Beiersdorfer:1996:4000} to search the feature from the $1s2p \, ^{3\!}P_{2} \rightarrow 1s2s \, ^{3\!}S_{1}$ transition in U$^{90+}$, were employed also to independently measure the line of interest in U$^{89+}$, which resulted in the value $4459.55$~eV. The uncertainty for this measurement was not specified in Ref.~\cite{Beiersdorfer:1996:4000}, but it seems reasonable to assume that it is comparable in magnitude with the typical errors reported in Refs.~\cite{Beiersdorfer:1993:3939, Beiersdorfer:1995:114, Beiersdorfer:1996:4000} providing even better agreement with our theoretical prediction $4459.57(10)$~eV for Li-like uranium.

\begin{table}[t]
\centering

\renewcommand{\arraystretch}{1.25}

\caption{\label{tab:relative:comparison} 
         Energy differences between the $2p_{3/2} \rightarrow 2s$ transitions in He- and Li-, He- and Be-, and Be- and Li-like uranium (in eV). See the text for details.
         }

\begin{tabular}{
                l@{}
                l@{\quad}
                S[table-format=2.3(2)]                
                S[table-format=2.3(2)] 
                S[table-format=2.3(2), table-align-text-post=false]                     
               }
               
\hline
\hline

   \multicolumn{2}{c}{\rule{0pt}{1.2em}Contribution}                                &
   \multicolumn{1}{c}{\parbox{1.8cm}{\centering $\text{He}-\text{Li}$~}}            &
   \multicolumn{1}{c}{\parbox{1.8cm}{\centering $\text{He}-\text{Be}$~}}            &
   \multicolumn{1}{c}{\parbox{1.8cm}{\centering $\text{Be}-\text{Li}$~}}            \\     
        
\hline   

 $E_{\text{non-QED}}$ &                                 &  51.526                                                   &  10.607      &  40.919      \\
 
 $E_{\rm QED}$        &                                 &  -1.216                                                   &  -2.502      &   1.286      \\
 
                      &  $E^{\rm QED}_{\text{1e-1l}}$   & \multicolumn{1}{c}{\parbox{1.8cm}{\centering ---~~~~~~}}  &  -1.079      &   1.079      \\
                      
                      &  $E^{\rm QED}_{\text{1e-2l}}$   & \multicolumn{1}{c}{\parbox{1.8cm}{\centering ---~~~~~~}}  &   0.007      &  -0.007      \\
                      
                      &  $E^{\rm QED}_{\rm many}$       &  -1.216                                                   &  -1.428      &   0.212      \\
                      
                      &  $E^{\rm QED}_{\rm rec}$        &   0.000                                                   &  -0.001      &   0.001      \\
                      
 $E_{\rm tot}$        &                                 &  50.311(39)                                               &   8.106(43)  &  42.205(39)  \\
 
 \multicolumn{2}{l}{Expt. \cite{Trassinelli:2009:63001, Trassinelli:2011:014003}}  &  50.34(96)  &    &                 \\
 
 \multicolumn{2}{l}{Expt. \cite{Beiersdorfer:1993:3939, Beiersdorfer:1995:114}}    &             &    &  42.35(48)$^a$  \\ 

\hline 

\end{tabular}%

{
\begin{flushleft}
$^{a}$ The value is obtained as the differences of the measurements for U$^{88+}$ and U$^{89+}$. The uncertainties are added linearly.
\end{flushleft}
}

\end{table}

Finally, Table~\ref{tab:relative:comparison} presents our results for the differences between the $2p_{3/2} \rightarrow 2s$ transition energies in He-, Li-, and Be-like uranium. The non-QED and QED contributions are shown separately. As noted above, the non-QED term is obtained by adding the frequency-dependent correction of the one-photon-exchange contribution, the non-QED part of the nuclear recoil effect, and the nuclear polarization and deformation corrections to the Breit-approximation value resulting from the DCB equation. The QED term covers the remainder. For better representation of the results, we additionally divide the QED contribution into the one-electron one-loop term, $E^{\rm QED}_{\text{1e-1l}}$, the one-electron two-loop term, $E^{\rm QED}_{\text{1e-2l}}$, the many-electron QED term, $E^{\rm QED}_{\rm many}$, and the QED part of the nuclear recoil effect, $E^{\rm QED}_{\rm rec}$. The term~$E^{\rm QED}_{\rm many}$ consists of the two-electron one-loop contribution, $E^{\rm QED}_{\text{2e-2l}}$, the part of the two-photon-exchange contribution beyond the Breit approximation, $E^{\rm QED}_{\rm 2ph}$, and the higher-order screened QED correction, $E^{\rm QED}_{\rm h.o.}$. Keeping in mind that the individual contributions for Be-like uranium are not exactly additive, we note that the values, shown in Table~\ref{tab:relative:comparison} for the ``$\text{He}-\text{Be}$'' and ``$\text{Be}-\text{Li}$'' differences, have been evaluated in accordance with the results obtained in Table~\ref{tab:Be:details}. The consideration of the three charge states within the single calculation has made it possible to obtain the most precise theoretical predictions for their pairwise differences. Due to the cancellations of the one-electron contributions, the uncertainties for them are significantly reduced compared to the ones for the transition energies. For the ``$\text{He}-\text{Li}$'' difference, as discussed above, the one-electron QED contributions vanish completely. The term~$E^{\rm QED}_{\rm rec}$ turns out to be zero, since within the independent-electron approximation not only the one-electron QED recoil contributions, but also the two-electron ones coincide for the $2p_{3/2} \rightarrow 2s$ transitions in He- and Li-like uranium~\cite{Malyshev:2018:085001}. The interelectronic-interaction correction to the QED nuclear recoil effect would result in a nonzero value, which we believe will be within the limits of the estimate made. For the ``$\text{He}-\text{Be}$'' and ``$\text{Be}-\text{Li}$'' differences, the one-electron contributions are reduced but nonzero due to the strong level mixing. Good agreement with the experimental value for the ``$\text{He}-\text{Li}$'' difference from Refs.~\cite{Trassinelli:2009:63001, Trassinelli:2011:014003} is found. In addition, we have evaluated the difference of the $2p_{3/2} \rightarrow 2s$ transition energies for Be- and Li-like uranium measured on the EBIT in Refs.~\cite{Beiersdorfer:1993:3939, Beiersdorfer:1995:114}. It also agrees well with our prediction. Nevertheless, the further improvement of the experimental accuracy and the extension of the analysis of Refs.~\cite{Trassinelli:2009:63001, Trassinelli:2009:012026, Trassinelli:2011:014003} to the case of the ``$\text{He}-\text{Be}$'' and ``$\text{Be}-\text{Li}$'' differences are in demand. As it is seen from Table~\ref{tab:relative:comparison}, the study of the ``$\text{He}-\text{Be}$'' and ``$\text{He}-\text{Li}$'' differences can give a test of the many-electron QED effects in the strongest Coulomb field at the 3\% level, provided the required experimental accuracy is achieved.


\section{Summary \label{sec:3}}

In the present work, the $2p_{3/2} \rightarrow 2s$ transition energies in He-, Li-, and Be-like uranium and their pairwise differences have been rigorously evaluated within the \textit{ab initio} QED approach. The calculations are carried out in the framework of the Furry picture as well as its generalizations, which imply the inclusion of some local screening potentials into the zeroth-order Hamiltonian. Four different screening potentials have been employed for this aim. The contributions of all first- and second-order Feynman diagrams are taken into account within QED perturbation theory for a single level in the case of U$^{90+}$ and U$^{89+}$ and its counterpart for quasidegenerate levels in the case of U$^{88+}$. The third- and higher-order electron-correlation effects are treated in the Breit approximation. The higher-order screened QED effects are estimated within the model-QED-operator approach. The contributions of the nuclear recoil, polarization, and deformation effects are also taken into account. All the possible sources of the uncertainties have been diligently studied. As a result, the most accurate theoretical predictions for the transitions themselves and for the transition-energy differences have been obtained. The extensive comparison with the previous relativistic calculations and experimental data is done. Perfect agreement with the results of the available measurements is found. The present calculations may serve as the benchmarks ones for the future high-precision experiments to measure the intra-shell transitions in highly charged uranium. 


\section*{Acknowledgments}

We thank M. Trassinelli for drawing our attention to the issue. The work was supported by the Russian Science Foundation (Grant No. 22-62-00004, https://rscf.ru/project/22-62-00004/). The calculations of the two-photon-exchange diagrams were supported by the Foundation for the Advancement of Theoretical Physics and Mathematics BASIS (Project No. 21-1-3-52-1).




\end{document}